\documentclass[10pt,preprint]{sigplanconf}
\usepackage{multirow}
\usepackage{amsmath}
\usepackage{amssymb}
\usepackage{empheq}
\usepackage{algorithm}
\usepackage{algorithmicx}
\usepackage[noend]{algpseudocode}
\usepackage[vlined,linesnumbered,ruled,algo2e]{algorithm2e}
\usepackage{color}
\usepackage{booktabs}
\usepackage{hyperref}
\usepackage[normalem]{ulem}
\usepackage{verbatim}
\usepackage{listings}
\usepackage[labelformat=simple]{subcaption}

\usepackage{cancel}
 
\usepackage{setspace}
\DeclareMathAlphabet\mathbfcal{OMS}{cmsy}{b}{n}
\definecolor{dmlgreen}    {RGB}{51,  160,  44}
\definecolor{dmlblue}     {RGB}{31,  120, 180}
\definecolor{dmlred}      {RGB}{202,   0,  32}
\usepackage{listings}

\newcommand{\ttt}{\texttt}

\begin{document}
\title{Expressing Sparse Matrix Computations for Productive Performance on Spatial Architectures}%\vspace{-0.6em}}

\authorinfo{Hongbo Rong}{Parallel Computing Lab (PCL), Intel}{hongbo.rong@intel.com}

\maketitle

\begin{abstract}
This paper addresses spatial programming of sparse matrix computations for productive performance. {\it The challenge is how to  express an irregular computation and its optimizations in a regular way}.

A sparse matrix has (non-zero) values and a structure. In this paper, we propose to classify the implementations of a computation on a sparse matrix into two categories: (1) structure-driven, or top-down, approach, which traverses the structure with given row and column indices and locates the corresponding values, and (2) values-driven, or bottom-up, approach, which loads and processes the values in parallel streams, and  decodes the structure for the values' corresponding row and column indices.

 On a spatial architecture like FPGAs, the values-driven approach is the norm. We show how to express a sparse matrix computation and its optimizations for a values-driven implementation. A compiler automatically synthesizes a code to decode the structure. In this way, programmers focus on optimizing the processing of the values, using familiar optimizations for dense matrices, while leaving the complex, irregular structure traversal to an automatic compiler. We also attempt to regularize the optimizations of the reduction for a dynamic number of values, which is common in a sparse matrix computation.

\end{abstract}

\section{Introduction}
\label{sec:intro}

This paper addresses high-performance high-productivity programming of sparse matrix computations on spatial architectures like FPGAs. Such computations are usually considered irregular in terms of memory access patterns and parallelism, etc. For productive performance, however, we should hide the irregularity and express the computations and their optimizations in a regular way. How to do so remains an open problem. There have been some efforts on CPUs~\cite{Arnold:2010:SVS:1863543.1863581,Kjolstad:2017:TAC:3152284.3133901}, but no work has ever been done on spatial architectures, as far as we know. 

A {\it sparse matrix} is a 2-dimensional storage with a lot of zeros in it. This storage is a unique {\it dense representation}, where every datum, zero or not, is stored, and is located by a unique row and column index.

However, a sparse matrix can have many sparse representations. A {\it sparse representation} has {\it values} and a {\it structure}, where the values include all the non-zeros, and the structure is a mapping from the row and column indices to the values. Popular sparse representations include CSR (Compressed Sparse Row), CSC (Compressed Sparse Column), and their blocked versions, etc~\cite{wiki:sparse:matrix,Arnold:2010:SVS:1863543.1863581}.

\begin{figure*}[tb]
\centering
\hspace*{\fill}
\subcaptionbox{Dense representation. Zeros not shown.\label{fig:trivial-sparse-matrix}}{%
  \includegraphics[scale=.75]{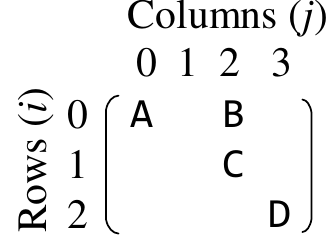}}\hfill
\subcaptionbox{A sparse representation in a tree.\label{fig:trivial-sparse-matrix-TACO}}{%
\includegraphics[scale=0.75]{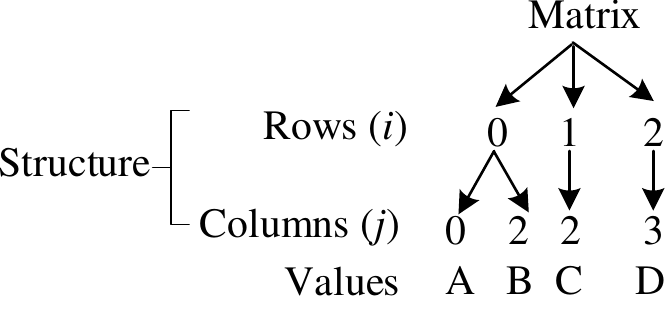}}\hfill
\subcaptionbox{The equivalent CSR representation.\label{fig:trival-sparse-matrix-csr}}{%
\includegraphics[scale=0.75]{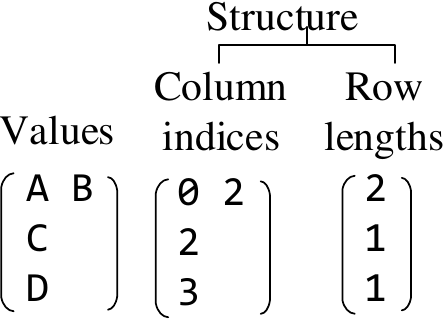}}\hfill
\hspace*{\fill}%
%\vspace{-1em}
\caption{A sparse matrix's representations.}~\label{fig:trivial-sparse-matrix-reps}
\\
\centering
\hspace*{\fill}
\includegraphics[scale=.75]{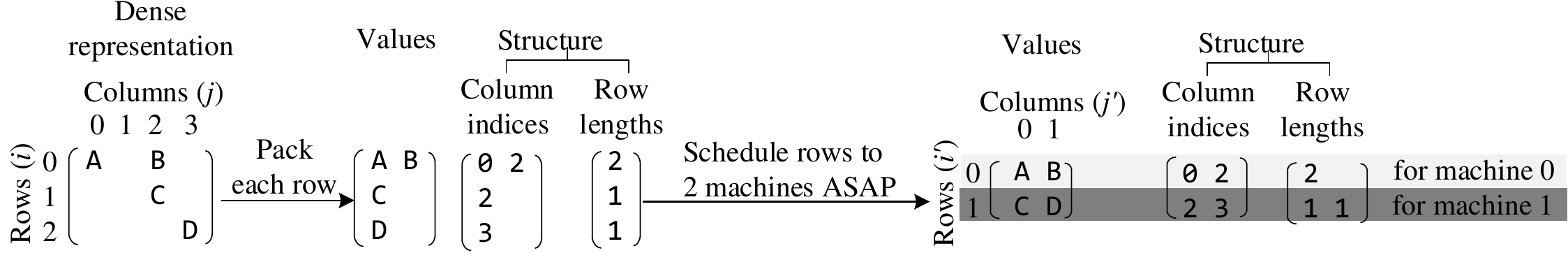}\hfill
\caption{Illustration: Evolving the representation of a sparse matrix.}~\label{fig:trivial-illustration}
\end{figure*}

Fig.~\ref{fig:trivial-sparse-matrix-reps} illustrates a few representations of a sparse matrix. Fig.~\ref{fig:trivial-sparse-matrix-reps}(a) shows its dense representation. Fig.~\ref{fig:trivial-sparse-matrix-reps}(b) is a sparse representation in a tree, where every dimension corresponds to a tree level, and the values are located at the leaf nodes. This is the way proposed in The Tensor Algebra Compiler (TACO)~\cite{Kjolstad:2017:TAC:3152284.3133901}~\footnote{In TACO, a tree is used to conceptually illustrate how a tensor can be represented. The physical representation is in dense vectors. A tensor is a generalization of a matrix to higher dimensions.}. This tree representation is equivalent to a CSR representation(Fig.~\ref{fig:trivial-sparse-matrix-reps}(c)). 

In this paper, we propose to classify the implementations of a computation on a sparse matrix into two categories:
\begin{itemize}
\item {\bf Structure-driven, or top-down, approach}, which traverses the structure with given row and column indices, and locates the corresponding values.
    
Take sparse-matrix dense-vector multiply (SpMV), an important kernel in sparse linear algebra, for example:
\begin{equation}
 \texttt{y(i)} = \sum_{{\forall \ttt{j}, } \texttt{A(i, j)} \ne 0}\texttt{A(i, j)} \cdot \texttt{x(j)},
 \label{equ:spmv-index-driven}
\end{equation}
Here {\ttt y} and {\ttt x} are dense column vectors, {\ttt A} is a sparse  matrix, and  {\ttt A(i, j)} refers to the value of matrix {\ttt A} with  row index {\ttt i} and column index {\ttt j}. In order to realize the computation, the structure of matrix {\ttt A} is traversed, and the corresponding values are located and loaded. 

For the tree representation in Fig.~\ref{fig:trivial-sparse-matrix-reps}(b), the tree is traversed top-down until a value is reached at the bottom. This structure-driven approach is commonly seen in CPU implementations~\cite{Kjolstad:2017:TAC:3152284.3133901} and sometimes on GPUs~\cite{Deveci:spmm:journals/corr/abs-1801-03065}. 

\item {\bf Values-driven, or bottom-up, approach}, which loads and processes the values in parallel streams, aiming to maximize memory bandwidth, data locality and parallelism. The structure is decoded to locate the indices of the values~\cite{Fowers:2014,Zhuo:2005:SMM:1046192.1046202,Greathouse:Daga:SpMV:2014,Bell:multigrid:2012,Bell:SpMV:GPU:2009:ISM:1654059.1654078}. 

The decoding of the structure is bottom-up: if we take the tree representation in Fig.~\ref{fig:trivial-sparse-matrix-reps}(b) again, for a value, we need traverse the tree bottom-up so as to determine the indices of a value. 

This values-driven approach particularly suits a spatial architecture: the key to get high performance there is to stream values in and consume the values in massive parallelism. It would be cumbersome if a structure has to be decoded first in order to load values.
\end{itemize}

For our purpose of high-performance high-productivity spatial programming, we need productively express values-driven implementations. As we said, a sparse matrix computation like that in Equation~\ref{equ:spmv-index-driven} is suitable for a structure-driven implementation. How could we re-express the computation to make it amenable to a values-driven  implementation, instead?

We observe that {\it a sparse matrix can be represented by a series of invertible transformations, including but not limited to, packing, blocking, and job scheduling}. Packing puts the non-zeros of the matrix together along the rows or columns of the matrix. Blocking divides the non-zeros into blocks. Job scheduling schedules the non-zeros, in rows, columns, or blocks, as jobs to some ``machines'', which are hardware units with identical logic and will process the non-zeros in parallel.

These transformations must be invertible: their resulting structure must encode the original row and column indices in some way so that the one-to-one correspondence between a value and its indices is still maintained.

Fig.~\ref{fig:trivial-illustration} illustrates the observation with the same sparse matrix in Fig.~\ref{fig:trivial-sparse-matrix-reps}. Starting from the dense representation, first pack the non-zeros in each row together, and we get the CSR sparse representation. Then we schedule the rows to 2 ``machines'' in ASAP (As Soon As Possible) policy, resulting another representation called CISR (Condensed Interleaved Sparse Representation). Because the job scheduling scheduled the jobs to 2 machines, in this final representation, there are 2 rows of values. The first (second) row of the values, along with related structure information, will be fed to the first (second) machine. Each machine accepts its inputs as a stream and processes the stream independently. This is essentially the SpMV design proposed by Fowers, et al.~\cite{Fowers:2014}, where the machines processes the streams by multiply-add. 

This is the first time that job scheduling is proposed as a primitive transformation for representing a sparse matrix: since we target a spatial architecture that allow many hardware units (i.e. machines) with the exactly the same circuit to perform computation in parallel, scheduling values to the machines is a natural abstraction. 

The transformations can be applied more than once, but the last transformation should always be a job scheduling, for the same reason stated above. This last job scheduling gives a value two new indices: a machine identifer ${\ttt i^\prime}$, and the position ${\ttt j^\prime}$ of a value within the stream scheduled to that machine. 

From now on, we will use {\ttt i} and {\ttt j} to represent the row and column index of a value in the dense representation, respectively, and ${\ttt i^\prime}$ and ${\ttt j^\prime}$ to represent the new indices in the final sparse representation, respectively. The two pairs of indices are 1-1 corresponding to each other. We call the mapping from ({\ttt i}, {\ttt j}) to  (${\ttt i^\prime}$, ${\ttt j^\prime}$) the {\it forward mapping}, and the mapping from (${\ttt i^\prime}$, ${\ttt j^\prime}$)  to  ({\ttt i}, {\ttt j}) the {\it inverse mapping}.

With the inverse mapping, we can re-express a sparse matrix computation, e.g. SpMV in Equation~\ref{equ:spmv-index-driven}, as follows:
\begin{equation}
 {\ttt y}(({\ttt i}^\prime, {\ttt j}^\prime) \longrightarrow {\ttt i}) = 
 \sum_{\forall {\ttt i}^\prime, {\ttt j}^\prime} \texttt{A.values}({\ttt i}^\prime, {\ttt j}^\prime) \cdot \texttt{x}(({\ttt i}^\prime, {\ttt j}^\prime) \longrightarrow {\ttt j}).
 \label{equ:spmv-values-driven}
\end{equation}
Here $\texttt{A.values}({\ttt i}^\prime, {\ttt j}^\prime)$ represents the ${\ttt j}^\prime$'th value to be processed by machine ${\ttt i}^\prime$,  and $({\ttt i}^\prime, {\ttt j}^\prime) \longrightarrow {\ttt i}$ and $({\ttt i}^\prime, {\ttt j}^\prime) \longrightarrow {\ttt j}$ are the inverse mapping. For each machine ${\ttt i}^\prime$, we may increment ${\ttt j}^\prime$ so that effectively, each machine is loading a stream of data, {\it without} accessing the structure of the matrix. This is exactly what a high-performance implementation on a spatial hardware would behave. Therefore, the above equation enables a computation to be driven by values.

Remember that \texttt{A.values}, {\ttt x} and {\ttt y} are all densely stored. Thus one may apply the familiar dense-matrix/vector optimizations, such as tiling, unrolling, etc., in optimizing the computation in Equation~\ref{equ:spmv-values-driven}.

The inverse mapping, $({\ttt i}^\prime, {\ttt j}^\prime) \longrightarrow {\ttt i}$ and $({\ttt i}^\prime, {\ttt j}^\prime) \longrightarrow {\ttt j}$, may be automatically generated by a compiler by inverting the high-level transformations used for a sparse representation like packing, etc. Usually, a traversal of a structure for either the forward mapping or the inverse mapping would require code to be written in an imperative language with complex array indirections and control flow, which is especially complicated when processing many values in parallel. The diverse representations a sparse matrix might have further increase the complexity. Now that the complex, irregular structure traversal is automatically handled by a compiler, programmers can be more productive and focus on optimizing the processing of values instead. 

In this paper, we further propose how to regularize and optimize a reduction with dynamic number of values, which is common in a sparse matrix computation. We show how to partition a reduction into two levels of reductions, and specify useful properties that help a compiler to generate efficient reduction circuits.

\begin{figure*}[tb]
\centering
\includegraphics[scale=.75]{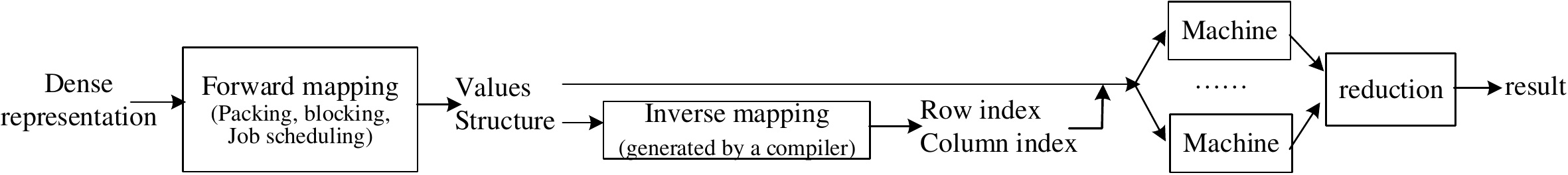}
\caption{Our approach for high-performance sparse matrix computation.}~\label{fig:overall}
\end{figure*}

Fig.~\ref{fig:overall} summarizes our approach how to, in general, efficiently implement a sparse matrix computation on a spatial hardware. Starting from the dense representation of the sparse matrix, a programmer specifies a sparse representation as a forward mapping, using a series of high-level transformations, including packing, blocking, and job scheduling, etc. A compiler automatically constructs an inverse mapping. The programmer specifies the optimizations of the matrix so that the values and the inverse mapping are fed into the machines that the last job scheduling targeted. Between the machines, there might be interactions, for example, reduction.
Finally, the program specifies the results of the computation to be stored back to the external memory of the spatial architecture.  

Our language is named T2S (Temporal To Spatial), which is an extension of the Halide language~\cite{Ragan-Kelley:2013:HLC:2491956.2462176} to spatial architectures. Previously, we have proposed T2S in the context of dense matrices~\cite{Rong:2018:T2S:arxiv}, where a sparse matrix cannot be directly represented unless it is pre-processed into a dense matrix on the host, and for an inverse-mapping, an external function written in an imperative language has to be introduced.

In this paper, we show how to further extend T2S to express and process sparse matrices directly. To the best of our knowledge, this is the first time a sparse computation can be specified in a productivity programming language so that its optimizations can also be specified in a way much like for a dense matrix computation.

Below we introduce the language extension of T2S for sparse matrices, and illustrate the approach with two high-performance SpMV designs on FPGAs. We skip a detailed description of the T2S language but introduce it briefly when explaining the examples. Then we discuss related work, point out  future directions, and conclude the paper.

\section{Language}
\label{sec:language}

We extend T2S in two aspects for sparse matrices. One aspect is how to specify forward and inverse mapping, the other is how to specify optimizations of reductions. The other optimizations like tiling, unrolling, buffering, etc. are the same as for dense matrices. 

\subsection{Forward and Inverse Mapping}
\label{subsec:mapping}

We explicitly express a sparse matrix as the result of a series of invertible transformations. For example, the following specification expresses an input matrix {\ttt A} in CSR sparse representation:
\begin{lstlisting}
    1 Representation CSR;
    2 Parameter      A(float, 2, CSR);
    3 Var            col_idx;
    4 CSR = A.dense()
    5        .pack(Columns, Index(col_idx));
\end{lstlisting}

Here Line 1 declares a representation named CSR, but exactly what it is is not defined yet. Line 2 says that input parameter {\ttt A} is a 2-dimensional floating-point matrix in CSR representation. Line 3 declares a variable {\ttt col$\_$idx}, which will be used to refer to the column indices in the CSR representation. Then we define CSR: starting from the matrix's dense representation (Line 4), pack the non-zeros along the column dimension (Line 5). The packed dimension is represented sparsely based on indices,  referred to by the pre-declared variable {\ttt col$\_$idx}, which gives us a handle to access the indices later.

As illustrated in Fig. 1, the sparse representation of a sparse matrix can continuously evolve. Continue with the above example. There can be a subsequent statement

\begin{lstlisting}
    6 Representation CISR = A.schedule(Rows, 4,
                                   ASAP, Zero);
\end{lstlisting}

This statement further schedules the rows of matrix {\ttt A} to 4 machines in the ASAP policy, and pads the machines with zeros at the end if needed, so that every machine has the same number of values to process. At this moment, all we need is only the total number of machines. Exactly how a machine behaves can be defined later. 

Below we define the language elements:

\begin{itemize} 
\item {\ttfamily Representation r} \\
Declare a representation named as {\ttt r}.

\item {\ttfamily Parameter matrix(type, total dimensions, r)} \\
Declare an input matrix with the given type, number of dimensions, and representation {\ttt r}.

\item {\ttfamily	matrix.dense()}\\
The dense representation of the sparse matrix.

\item {\ttfamily	matrix.pack(d, r)}\\
Pack the matrix along dimension d, which is either {\ttt Rows} or {\ttt Columns}. The resulting representation {\ttt r} can be index-based, i.e. {\ttfamily Index(idx)}, or with a length as well, i.e. {\ttfamily Index(idx), Length(len)}, where {\ttt idx} and {\tt len} are variables recording the indices and length of the values along the packed dimension.

\item	{\ttfamily matrix.block(d, f, padding policy, r)}\\
Block dimension {\ttt d} with factor {\ttt f}, with the given padding policy. The resulting representation {\ttt r} is {\ttfamily Length(b)}, where {\ttt b} is a variable for the number of blocks along the dimension.

\item {\ttfamily matrix.schedule(d, m, scheduling policy, padding policy)}\\
Schedule dimension {\ttt d} to {\ttt m} number of machines, with a given scheduling policy. At the end, balance the machines' workloads in a given padding policy.

\item {\ttfamily matrix.values(i$^\prime$, j$^\prime$)}\\
After the above job scheduling, access one value of the sparse matrix by the machine identifier {\ttt i$^\prime$} and the position {\ttt j$^\prime$} of the value within the stream scheduled to machine {\ttt i$^\prime$}.

\item {\ttfamily matrix.inverse$\_$map(i$^\prime$, j$^\prime$)}\\
After the above job scheduling, for a value scheduled to machine {\ttt i$^\prime$} at position {\ttt j$^\prime$}, return an expression that represent the row and column index in the corresponding dense representation of the sparse matrix. Let that expression be {\ttt e}, we can extract the row and column index by {\ttt e[0]} and {\ttt e[1]}, respectively.

\item {\ttfamily Func.load$\_$structure(var1, var2, ...)}\\
Load the structure information specifically referred to by {\ttfamily var1, var2}, ... These variables are those used in {\ttt pack} and {\ttt block} above.

\end{itemize}

\subsection{Reduction}
\label{subsec:reduction}

Reduction is a very common computation for sparse matrices. Unlike a reduction for a dense matrix or vector, one reduction for a sparse matrix usually has dynamic number of input values.

In this section, we will use the following conceptual example:

\begin{lstlisting}
    y((i$^\prime$, j$^\prime$) $\longrightarrow$ i) = initial value;
    y((i$^\prime$, j$^\prime$) $\longrightarrow$ i) += f(A.values(i$^\prime$, j$^\prime$));  
    \end{lstlisting}
where {\ttt f} is an arbitrary function. The above two definitions of {\ttt y}, respectively, are an {\it initial definition}, directly referred to as {\ttt y} in a specification, and an {\it update definition}, referred to as {\ttt y.update()} in a specification.

Usually, it is unknown to the compiler how many {\ttfamily (i$^\prime$, j$^\prime$)}'s are mapped to the same {\ttt i}, and whether these {\ttfamily (i$^\prime$, j$^\prime$)}'s appear next to each other to a machine. To help compiler generate  efficient reduction circuit, we can specify if the {\ttfamily (i$^\prime$, j$^\prime$)}'s that are mapped to the same {\ttt i} appear continuously to a machine, and the maximum number of {\ttfamily (i$^\prime$, j$^\prime$)}'s that are mapped to the same {\ttt i} to a machine. Further, if the number of {\ttfamily (i$^\prime$, j$^\prime$)}'s that are mapped to the same {\ttt i} must be a multiple of some integer number, we can block these {\ttfamily (i$^\prime$, j$^\prime$)}'s: reduce the corresponding values in each block to a local sum, then reduce the local sums of the blocks. This two-level reduction provides more optimization opportunities: the two reduction circuits can be optimized differently.

If several machines with different {\ttfamily i$^\prime$} reduce to the same {\ttt y(i)}, we may combine their reduction results and write into {\ttt y(i)} only once.

Below we describe the language features.

\begin{itemize} 
\item {\ttfamily f.is$\_$continuous$\_$reduction(l)}\\
For a specific iteration of loop {\ttt l} in function {\ttt f}, the function reduces to the same result location continuously before the function reduces to another result location.

\item {\ttfamily f.is$\_$distinct$\_$reduction(l)}\\
For two different iterations of loop {\ttt l} in function {\ttt f}, the function reduces to different result locations.

\item {\ttfamily f.isolate$\_$reduction(F)}\\
Isolate out of function {\ttt f} the reduction of values into a separate function {\ttt F}. This creates a two-level reduction: {\ttt F} reduces the values, and {\ttt f} reduces the results of {\ttt F}.

For the above example, we can isolate out a new reduction {\ttt Y} in the following specification:

\begin{lstlisting}
    Func   Y;
    y.update().isolate_reduction(Y);
\end{lstlisting}

This produces the following:

\begin{lstlisting}
    Y((i$^\prime$, j$^\prime$) $\longrightarrow$ i)  = 0;
    Y((i$^\prime$, j$^\prime$) $\longrightarrow$ i) += f(A.values(i$^\prime$, j$^\prime$));
    y(i)  = initial value;
    y(i) += Y(i);
\end{lstlisting}

Basically, \texttt{Y(i)} is a local sum. 
For the same {\ttt i}, {\ttt Y} might reduce several values and send the local sum to {\ttt y}, and then do the same for the next several values, and so on.

Every time {\ttt Y} needs to send {\ttt y} a local sum, it also sends the current {\ttt i} to tell exactly which {\ttt y(i)} to reduce this local sum for. The sending of {\ttt i} can be added automatically by a compiler.

\item {\ttfamily	f.combine$\_$reduction(l, target info)}\\
Combine all the reductions of all the iterations of loop {\ttt l} into a single reduction with additional target information, which can be {\ttfamily Same$\_$Target} or {\ttfamily Maybe$\_$different$\_$Targets}.

For the example at the beginning of Section~\ref{subsec:reduction}, we can combine the reductions of the iterations of loop \texttt{i$^\prime$} into a single reduction:

\begin{lstlisting}
    y.update().combine$\_$reduction(i$^\prime$,
                              Same$\_$Targets);
\end{lstlisting}

This transforms the original loop nest of the update definition of function {\ttt y}:
\begin{lstlisting}
    for j$^\prime$
     for i$^\prime$
      y((i$^\prime$, j$^\prime$) $\longrightarrow$ i) += f(A.values(i$^\prime$, j$^\prime$));
\end{lstlisting}

into the following:

\begin{lstlisting}
    for j$^\prime$
     sum = 0
     for i$^\prime$
       sum += f(A.values(i$^\prime$, j$^\prime$));
     y((0, j$^\prime$) $\longrightarrow$ i) = sum;
\end{lstlisting}

If, however, we specify
\begin{lstlisting}
    y.update().combine$\_$reduction(i$^\prime$,
                Maybe$\_$different$\_$Targets);
\end{lstlisting}

the original loop nest will be transformed into the following instead:

\begin{lstlisting}
    for j$^\prime$
     sum = 0
     (0, j$^\prime$) $\longrightarrow$  current$\_$i
     for i$^\prime$
       (i$^\prime$, j$^\prime$) $\longrightarrow$ i
       if (current$\_$i == i)
         sum += f(A.values(i$^\prime$, j$^\prime$));
       else
         y(current$\_$i) = sum;
         sum = f(A.values(i$^\prime$, j$^\prime$));
       current$\_$i = i
     y(current$\_$i) = sum;
\end{lstlisting}

\item {\ttfamily	f.reduction$\_$circuit(circuit structure, l)}\\
The reduction function {\ttt f} is implemented in a circuit with a given structure, including \texttt{Tree} and \texttt{Linear$\_$Array}. Usually, the reduction structure reduces the inputs level by level. The maximum number of levels in the circuit structure is {\ttt l}.

\end{itemize}

\section{Case Studies}
\label{sec:case_studies}

In this section, we illustrate our approach with two high-performance designs of SpMV on FPGAs. 

\subsection{SpMV on an FPGA with High Memory Bandwidth}
\label{subsec:spmv-ms-design}

In an implementation of SpMV on an FPGA~\cite{Fowers:2014} for the above Equation~\ref{equ:spmv-values-driven},  \texttt{j$^\prime$} is incremented by 1 every time such that the values for the same \texttt{i$^\prime$} are loaded from memory into an FPGA sequentially, forming a stream. And several streams for different \texttt{i$^\prime$}'s are formed in parallel. The final result, vector \texttt{y}, is stored back to the memory also as a stream. This implementation is focusing on the values, and can be very efficient for spatial architectures with high memory bandwidth.

\begin{figure*}[tb]
\centering
\includegraphics[scale=.8]{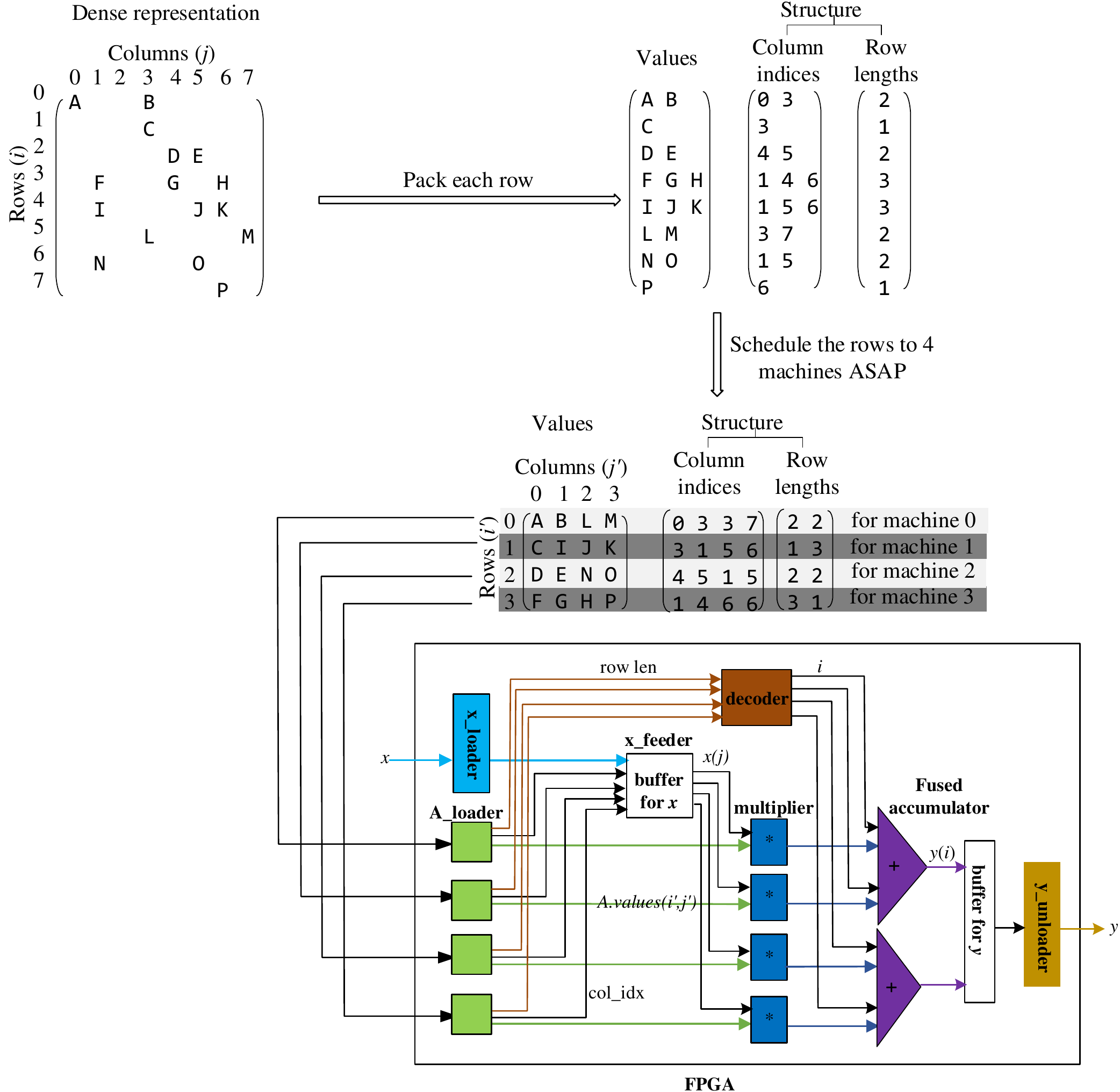}
\caption{Illustration: an SpMV design on an FPGA~\cite{Fowers:2014}. The example sparse matrix is  also from ~\cite{Fowers:2014}. }~\label{fig:spmv-ms-design}
\end{figure*}

Fig.~\ref{fig:spmv-ms-design} illustrates the implementation. Starting from the dense representation of a sparse matrix , pack together all non-zeros in each row, we get the CSR sparse representation. Let us say the target FPGA has 4 memory channels. We would like the matrix to be read from the 4 memory channels in parallel for the maximal memory bandwidth. If we treat the 4 memory channels as 4 ``machines'', and the compressed sparse rows as ``jobs'', we can schedule the jobs one by one to the machines: a job is scheduled to the earliest available machine; the last job for a machine may be padded with zeros so that all the machines have the same number of values to process. The resulting sparse presentation is CISR~\cite{Fowers:2014}. A similar but trivial example has been described in Section~\ref{sec:intro} before.

Now with the CISR sparse representation, every memory channel of the FPGA can read the rows scheduled to it as a sequential stream, and all the memory channels read their streams in parallel. More specifically, this is done by a hardware unit, i.e. 4 \texttt{A$\_$loader}s, in Fig.~\ref{fig:spmv-ms-design}. The loaded contents include row lengths, column index {\ttt j} and values.  A decoder figures out the row index {\ttt i} from the row lengths to each \texttt{A$\_$loader}. This decoder is realizing an inverse mapping \texttt{(i$^\prime$, j$^\prime$) $\longrightarrow$ i} based on the row lengths. Vector \texttt{x} and {\ttt y} are buffered on the FPGA. Vector \texttt{x} is loaded into its buffer with \texttt{x$\_$loader} in the figure. Using the row index {\ttt i} and column index {\ttt j}, \texttt{x(j)} and \texttt{y(i)} can be easily retrieved.  
Now all the necessary information are available for SpMV defined in Equation~\ref{equ:spmv-values-driven}. The computation is straightforward: 4 multipliers work in parallel to compute 4 different \texttt{A.values(i$^\prime$, j$^\prime$) * x(j)}. The products are reduced by, say 2, adders. Each adder processes two streams alternatively. Finally, the resulting vector \texttt{y} is stored back to memory by a hardware unit, \texttt{y$\_$unloader}, in the figure.

Note that the entire hardware in the figure is a regular dataflow graph. All the hardware units are purely functional. The only exception is the {\ttt decoder}, which is stateful and the output of it depends on its previous output.  Because it is not really functional, usually, one might have to write it in an imperative language, not a functional language. In our approach, however, a compiler should automatically generate such a decoder, i.e. generate the hardware for the inverse mapping \texttt{(i$^\prime$,  j$^\prime$) $\longrightarrow$ i}, automatically connecting the hardware with other hardware units in the figure based on the producer-consumer relationship.

\begin{figure*}[tb]
\centering
\includegraphics[scale=.8]{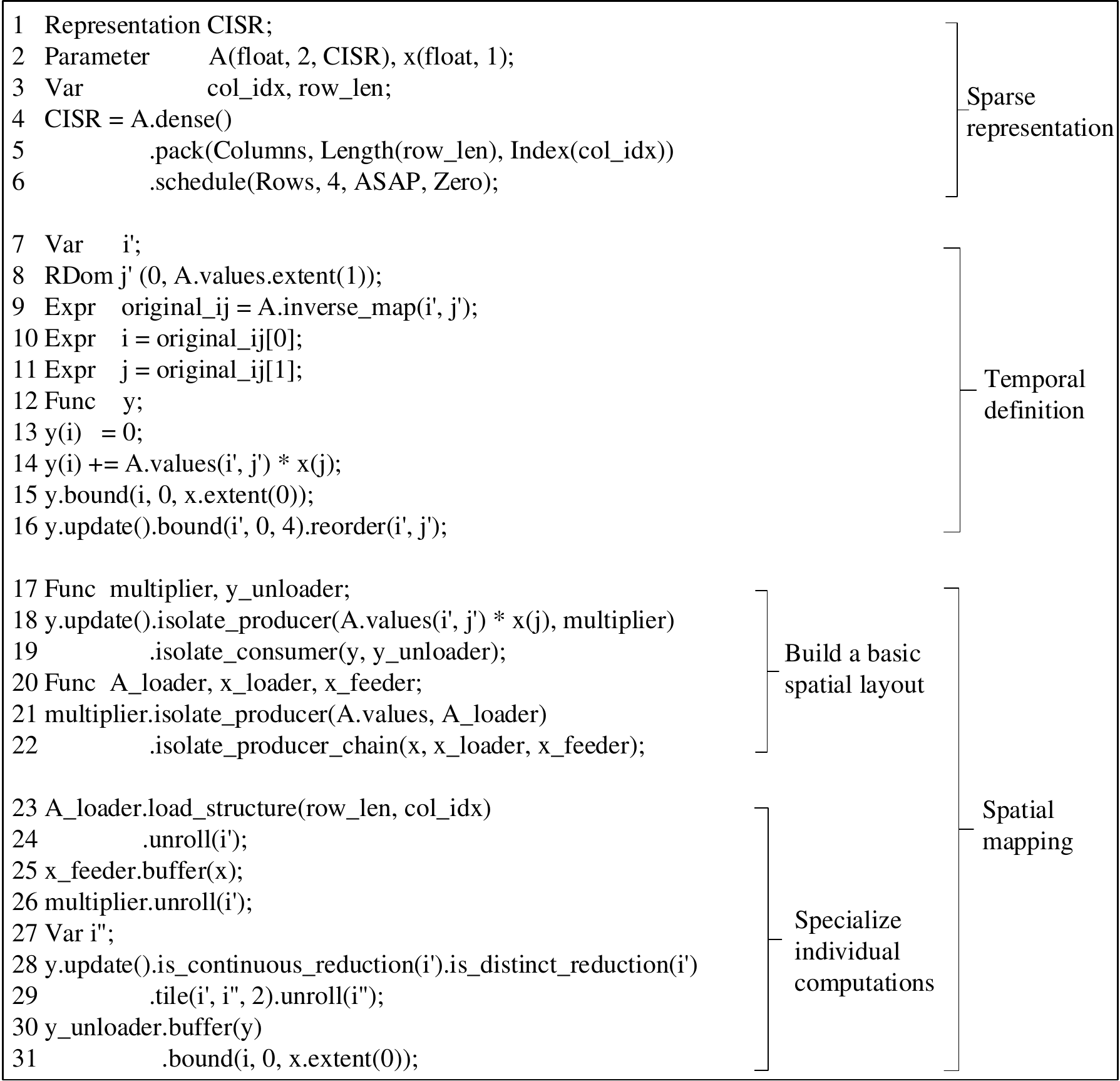}
\caption{Specification for the SpMV design in Fig.~\ref{fig:spmv-ms-design}.}~\label{fig:spmv-ms-spec}
\end{figure*}

Fig.~\ref{fig:spmv-ms-spec} shows the T2S specification. There are 3 parts: sparse representation, temporal definition, and spatial mapping. 

The sparse representation (Line 1-6) defines the sparse representation of the input matrix \texttt{A} as a series of transformations. Line 1 declares a sparse representation called CISR. Line 2 says that \texttt{A} is a 2-dimensional floating-point matrix with CISR sparse representation, and \texttt{x} is a floating-point vector. Line 3 declares two variables to be used next. Line 4-6 actually defines the sparse representation: starting from the dense representation of the sparse matrix, first, pack the nonzeros along the column dimension for each row, and record the total number and the column indices of the non-zeros in each row (\texttt{row$\_$len} and \texttt{col$\_$idx}). Then schedule the rows to 4 machines in ASAP policy with zero-padding at the end if needed.

The temporal definition (Line 7-16) defines what to compute. Line 10-11 defines the row and column index as two expressions based on an inverse mapping (Line 9). The machine identifier \texttt{i$^\prime$} is declared as normal variable (Line 7), while the position \texttt{j$^\prime$} of a value scheduled to a machine is a reduction variable whose range is from 0 to the last column of \texttt{A.values}. 

Line 14 defines the SpMV computation, according to Equation~\ref{equ:spmv-values-driven}. The result vector {\ttt y} is initialized as 0 beforehand in Line 13. We know that vector \texttt{y} is equally long as the input vector \texttt{x}. Thus instead of calculating every {\ttt i} from an inverse mapping, we can simply iterate {\ttt i} over the same range as vector \texttt{x} to initialize each \texttt{y(i)} directly (Line 15).

Line 16 says that the machine identifier \texttt{i$^\prime$} is from 0 to 4 (not included), as there are 4 machines. Further, the loops are ordered such that loop \texttt{i$^\prime$} is the inner loop, and loop \texttt{j$^\prime$} is the outer loop.

Line 17-22 isolate the computation in Line 14 into multiple pieces.  Line 18-19 first isolate the expression in the right-hand side of Line 14 into a function named \texttt{multiplier}, and then isolate the saving of the result vector \texttt{y} into another function \texttt{y$\_$unloader}. Further, Line 21 isolates from the multiplier the loading of the values of matrix \texttt{A} to a function \texttt{A$\_$loader}. Then Line 22 isolates from the multiplier the loading of vector \texttt{x} into a function \texttt{x$\_$loader}, which transfers the loaded values to another function \texttt{x$\_$feeder}, which transfers the values to function \texttt{multiplier}.

Line 23-31 specialize each function. In addition to loading the values of {\ttt A}, \texttt{A$\_$loader} is also specified to load the structure information \texttt{row$\_$len} and \texttt{col$\_$idx} (Line 23). Further, we unroll loop \texttt{i$^\prime$} in \texttt{A$\_$loader}, which creates 4 instances of \texttt{A$\_$loader} (Line 24). Line 25 lets \texttt{x$\_$feeder} buffers the  \texttt{x} values in an on-chip scratchpad memory, whose size will be automatically calculated by the compiler. 

Line 26 fully unrolls loop  \texttt{i$^\prime$} in \texttt{multiplier} to create 4 instances of it. Line 28 says that the reduction for {\ttt y} is continuous for each iteration of loop \texttt{i$^\prime$}, and is distinct for different iterations of loop \texttt{i$^\prime$}. Then Line 29 splits loop \texttt{i$^\prime$} by a factor of 2, and fully unrolls the newly created loop \texttt{i$^{\prime\prime}$} to creates two reduction circuits (i.e. the two fused accumulators in Fig.~\ref{fig:spmv-ms-design}). Line 30 buffers the entire result vector \texttt{y} before sending any data of it out. Line 31 says that in sending out the data of the buffered results of {\ttt y}, we can directly iterate {\ttt i} with the same range as the input vector \texttt{x}, instead of computing {\ttt i} indirectly from the inverse mapping \texttt{(i$^\prime$,  j$^\prime$) $\longrightarrow$ i}.  

The inverse mapping is specified as \texttt{A.inverse$\_$map()} in Line 9. The mapping should be realized by a compiler automatically to generate the decoder in Fig.~\ref{fig:spmv-ms-design}.
 
\subsection{Illustration: Another SpMV design on FPGAs~\cite{Zhuo:2005:SMM:1046192.1046202}}
\label{subsec:spmv-usc}

\begin{figure*}[tb]
\centering
\includegraphics[scale=.7]{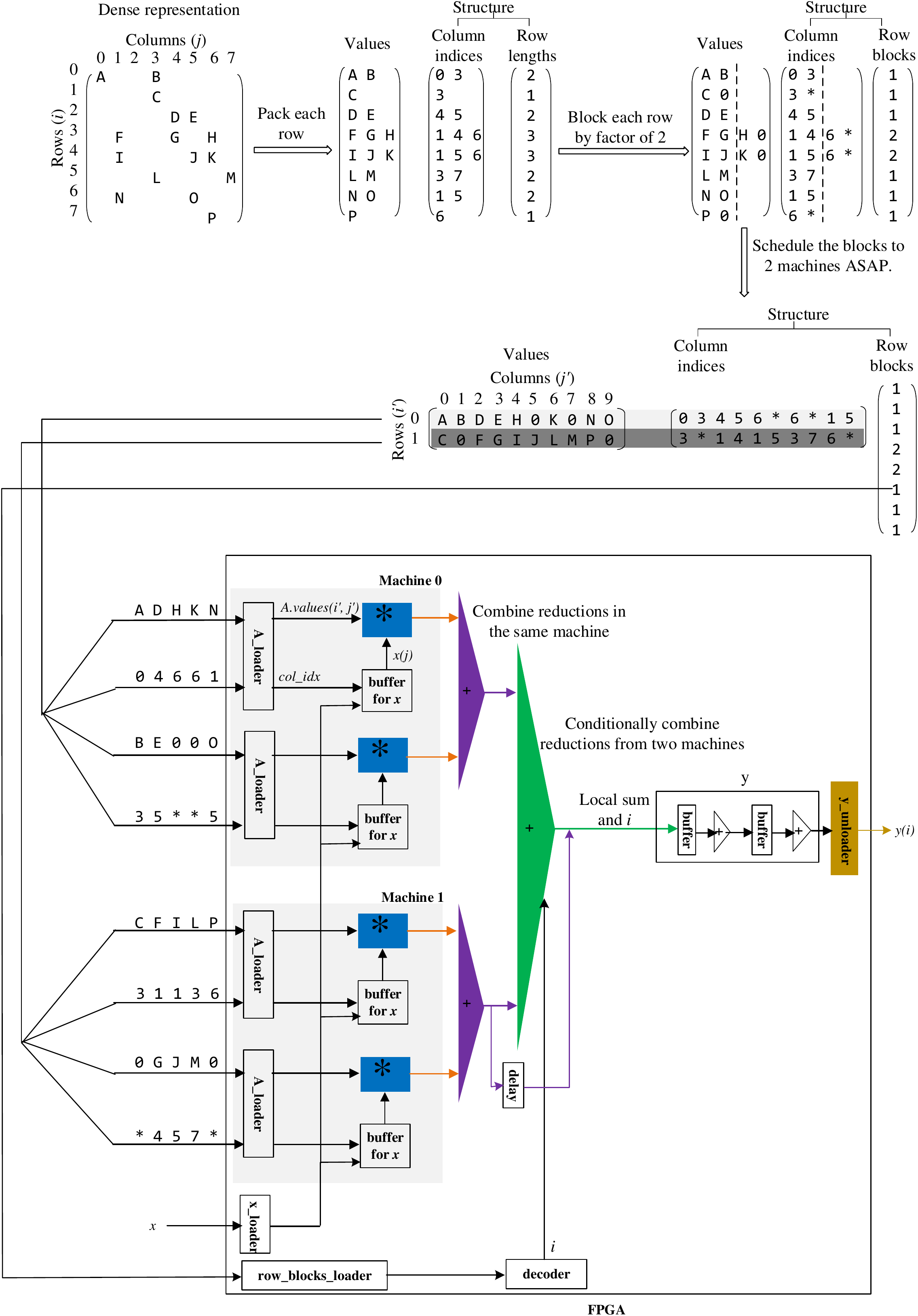}
\caption{Another SpMV design on FPGAs~\cite{Zhuo:2005:SMM:1046192.1046202}.}~\label{fig:spmv-usc-design}
\end{figure*}

This design is very different from the previous one in how the data are represented and processed.
Fig.~\ref{fig:spmv-usc-design} illustrates this design. After packing the columns along each row of the dense representation, we get the CSR sparse representation. Then with a constant {\ttt k} (e.g. 4), block each row by a factor of {\ttt k}/2 with zero-padding for the last block if needed. After blocking, a row is divided into blocks, and each block has {\ttt k}/2 values. Then schedule the blocks as jobs to 2 machines in the ASAP policy.

Each machine computes the sum of a block. For this purpose, it uses {\ttt k}/2 multipliers, each for one of the values in the block. Each multiplier gets the values and their column indices from another hardware unit,  \texttt{A$\_$loader}. With a column index, the multiplier reads a value of {\ttt x} from a buffer, which was filled beforehand by a hardware unit, \texttt{x$\_$loader}.  Now with a value from the matrix \texttt{A} and the corresponding value from the vector, the multiplier can compute their product. For all the multipliers in the same machine, their results are added together in a tree of adders (For the particular case shown in Fig.~\ref{fig:spmv-usc-design}, there is only 1 adder in the tree).

For the two machines, their corresponding blocks might be from the same row, or from two adjacent rows. A \texttt{row$\_$blocks$\_$loader} loads the structure information of the total blocks in each row, and sends the information to a {\ttt decoder}, which decides if the two blocks currently processed in the two machines belong to the same row or not. If yes, the {\ttt decoder} controls an adder to add together the two sums of the two blocks. Otherwise, the first sum is added with 0. The adder's result and the row index is sent to a reduction circuit. The second sum and the corresponding row index is also sent to the reduction circuit but with  some delay. The reduction circuit is accumulating the sums for all the blocks in the same row. It is composed of a linear arrays of small buffers and adders. With the row indices from the input, the reduction circuit knows when a row is completely reduced, and sends the results to another hardware unit, \texttt{y$\_$unloader}, to save to memory.

\begin{figure*}[tb]
\centering
\includegraphics[scale=.8]{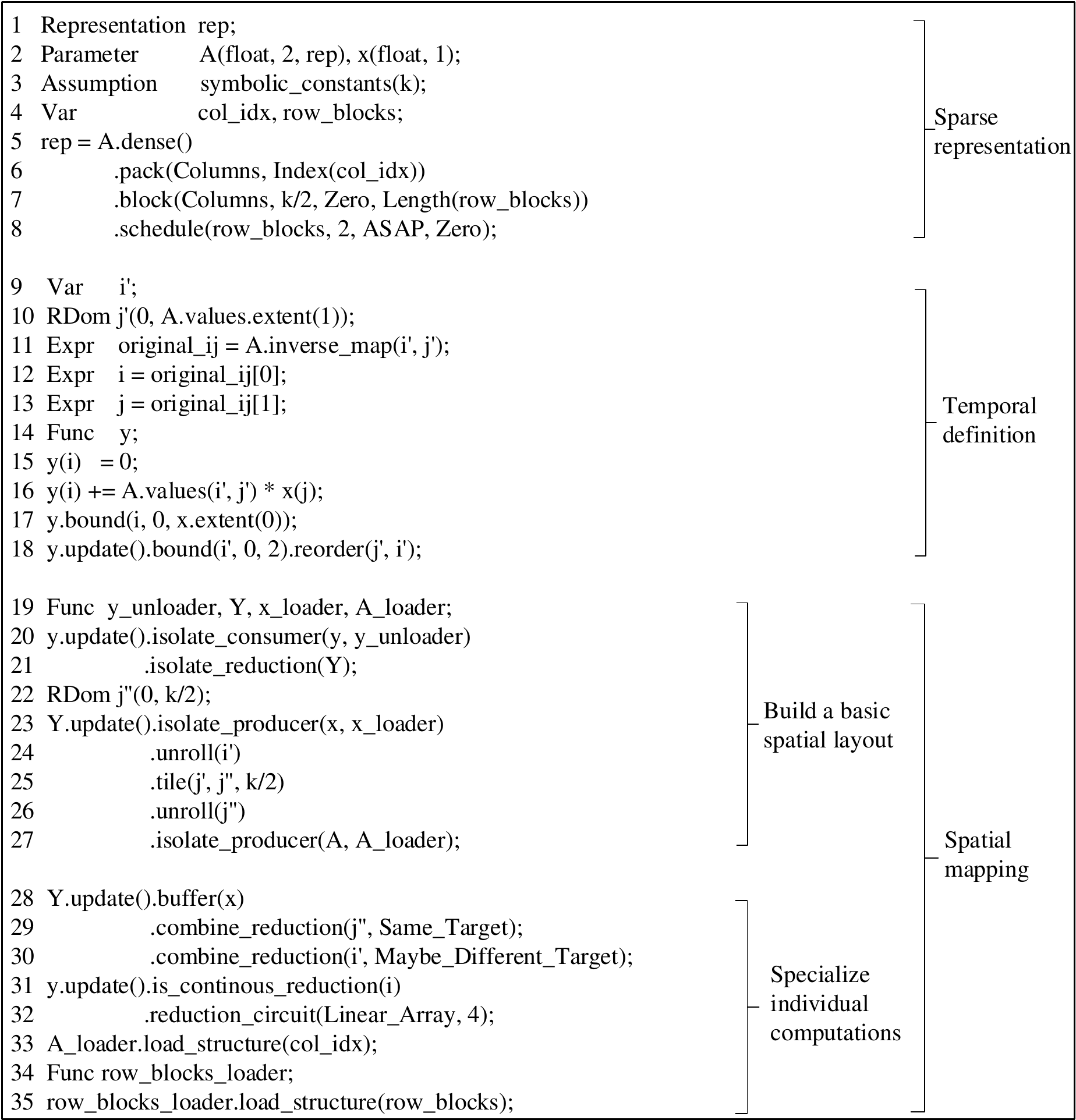}
\caption{Specification for the SpMV design in Fig.~\ref{fig:spmv-usc-design}.}~\label{fig:spmv-usc-spec}
\end{figure*}

Fig.~\ref{fig:spmv-usc-spec} shows the T2S specification for the design. Line 1-8 describe the sparse representation of the input matrix \texttt{A}. Line 2 says that A is a 2-dimensional floating-point sparse matrix in a representation called {\ttt rep}. How {\ttt rep} is defined? Starting from the dense representation of the matrix (Line 5), pack the non-zeros along the column dimension of each row and record the column indices of the non-zeros (\texttt{col$\_$idx}) (Line 6). Then block the columns of each row into blocks with a length of {\ttt k}/2 (Line 7), where {\ttt k} is a symbolic constant (Line 3),  and pad the last block with zeros if needed. The number of the blocks of a row is referred to by \texttt{row$\_$blocks}. Finally, schedules the blocks to 2 machines in ASAP policy, padding the last machine with a zeroed block if needed (Line 8).

Line 9-18 defines the temporal computation. Not surprisingly, it is the same as the temporal definition in the previous example in Fig.~\ref{fig:spmv-ms-spec}, except the number of machines are changed from 4 to 2,  the inner loop is \texttt{j$^\prime$} and the out loop is \texttt{i$^\prime$} (Line 18).

Line 19-27 builds a basic spatial layout. Line  20 isolates a \texttt{y$\_$unloader} to store the final results of vector \texttt{y} to the external memory. Line 21 isolates the reduction of {\ttt y} so that the reduction is done in two levels: there is an additional reduction circuit {\ttt Y} to compute local sums at the first level, which are further reduced at the second level. Line 23 isolates out the loading of vector \texttt{x}. Line 24 then unrolls loop \texttt{i$^\prime$} to create 2 machines. Line 25-26 split loop \texttt{j$^\prime$}  by a factor of {\ttt k}/2 and unroll the newly created loop \texttt{j$^{\prime\prime}$}   so that each machine has {\ttt k}/2 multipliers. Finally, Line 27 isolates the loading of matrix \texttt{A}.

Line 28-35 specialize the individual computations. Line 28 buffers vector \texttt{x} for {\ttt Y}. Then we combine the reductions in the same machine(Line 29).  We also combine the results of the 2 machines, but they may be reduce to two different result locations (Line 30).

Line 31 says that the local sums sent from function {\ttt Y} to function {\ttt y} must be continuous for a row. The reduction circuit for function {\ttt y} is chosen to be implemented in a linear array with 4 levels.

In addition to loading of {\ttt A} values, \texttt{A$\_$loader} also loads the structure information, \texttt{col$\_$idx} (Line 33).

Line 34 declares a special function, \texttt{row$\_$block$\_$loader}, which is neither defined nor isolated. This is because it is to fetch a part of the sparse structure information, the number of blocks per row. We simply specify that functionality in Line 35.

The inverse mapping is specified as \texttt{A.inverse$\_$map()} in Line 11. The mapping should be realized by a compiler automatically to generate the decoder in Fig.~\ref{fig:spmv-usc-design}.

\section{Related Work}
\label{sec:related}

We first discuss related work from the perspectives of implementing a sparse matrix computation, and then from the perspectives of languages.

\subsection{Structure-Driven vs. Values-Driven Approach} 
\label{subsec:structure-vs-values-driven}
 
In this paper, we classify algorithms for sparse matrix (or more generally, tensor) computations into two categories: the first is structure-driven, or top-down; the second is values-driven, or bottom-up.

Structure-driven algorithms are commonly seen for latency-oriented CPUs and sometimes for throughput-oriented GPUs. TACO~\cite{Kjolstad:2017:TAC:3152284.3133901} provides a unified and general way to represent all possible sparse tensors, including sparse matrices. Conceptually, every sparse tensor can be viewed as stored in a tree, with every tree level being a dense or sparse dimension of the tensor, and the leaves being the values of the tensor. The compiler will also automatically optimize the traversing of the structures of several related tensors. This traversal of the sparse structures is an efficient way of forward mapping, not the inverse mapping we propose in this paper. Thus the code generated by TACO is top-down, and is driven by structures.

In TACO, programmers control  the representations of a sparse tensor, but have no control of subsequent optimizations so far. In contrast, we would like programmers have explicit control not only in the representations of sparse matrices, but also how to optimize them. Also we target FPGAs, while TACO targets CPUs so far.

Gustavson~\cite{Gustavson:1978:TFA:355791.355796} proposed an algorithm for permuted transposition, and an algorithm for the multiplication of two sparse matrices (SpGEMM). Both algorithms follow the structure of a sparse matrix to get the values. Thus the algorithms are structure-driven. Based on Gustavson's algorithm for SpGEMM, Deveci et al.~\cite{Deveci:spmm:journals/corr/abs-1801-03065}  developed different parallel algorithms for portable performance accross different CPUs and GPUs. They consider the trade-offs on parallelism (how to partition and parallelize a reduction accross different compute units), how to determine the size of the result matrix, and efficient data structures of a reduction.     

Values-driven approach seems to be the norm for sparse matrix computation on FPGAs. This is not surprising: due to the massive hardware parallelism of FPGAs,  their slower frequencies relative to CPUs/GPUs, and their lack of hardware caches, a high-performance implementation on FPGAs would have to stream values in (maybe in several parallel streams) to avoid any random memory access, and process them in pipelines. Meanwhile, since FPGAs have abundant, programmable hardware resources, the implementation can use additional hardware resources to decode the structure of a sparse matrix for the indices of the values simultaneously.    

Values-driven approach also appears in GPU implementations. Greathouse and Daga~\cite{Greathouse:Daga:SpMV:2014} map SpMV efficiently to a GPU by streaming the values of a sparse matrix into the local scratchpad memory of the GPU, and dynamically assigning rows to each parallel GPU compute unit. Bell et al.~\cite{Bell:multigrid:2012} show an SpGEMM algorithm based on COO (coordinate) representation. Bell and Garland~\cite{Bell:SpMV:GPU:2009:ISM:1654059.1654078} compare several SpMV implementations based on different representations including DIA (diagonal), ELL, CSR and COO. While there are no detailed algorithm descriptions, we believe their algorithms are driven by values, because they target to distribute fine-grain parallelism among thousands or tens of thousands of threads, which might be difficult for a structure-driven approach. 

\subsection{Languages}
\label{subsec:languages}

Our language, T2S, is an extension of Halide~\cite{Ragan-Kelley:2013:HLC:2491956.2462176}, a domain-specific language for image processing on CPUs/GPUs, to spatial architectures. Halide-HLS~\cite{Pu:2017:Halide-HLS:3132652.3107953} is another spatial extension of Halide, where a dataflow graph of functions is specified to offload to an FPGA, with line buffers between functions to optimize their communication. As far as we know, sparse matrices are not expressible in the current Halide and Halide-HLS.

Spatial\cite{Koeplinger:2018:Spatial:3192366.3192379} is another domain-specific language for spatial architectures. Its syntax directly supports streaming, channels (i.e. pipes), reduction, finite state machine, loop parallelization, etc. There is no detail specific to sparse matrices, but it was evaluated with PageRank, a sparse matrix or graph algorithm. From its syntax, we believe that sparse matrices are not directly supported; instead, a values-driven approach might be implemented: the values of a sparse matrix may be streamed into a spatial hardware, and the structure of the matrix may be accessed with loops to build the inverse mapping. However, in this way, a programmer has to write in detail how to build the inverse mapping, while in our approach, we may have a compiler to automatically build it.    

The Little Language (LL)~\cite{Arnold:2010:SVS:1863543.1863581} has also observed that a sparse matrix can be represented by a series of transformations. However, LL does not point out that the transformations are invertible, a simple yet critical fact that makes it possible for a compiler to automatically construct an inverse mapping. The usage of the LL language was for verification of sparse matrix codes, which is structure-driven. In addition, the transformations in LL have much lower abstraction level. For example, blocking and job-scheduling here would have to be expressed in other lower-level primitives in LL. 

In our approach, we propose the transformations at a higher abstraction level in order to make it easy for a compiler to invert the transformations and automatically generate the inverse mapping, which is the key to enable a values-driven implementation.

\section{Future Work}
\label{sec:future}

This paper is a first step to expressing and optimizing sparse matrix computations in a values-driven style on spatial architectures. There are many interesting questions left for future:

\begin{itemize}
\item {\bf An algorithm for building an inverse mapping}. We reasonably assume that a compiler can automatically build an inverse mapping for a sparse matrix, as long as the matrix is represented in a series of invertible, high-level, transformations. In the next step, we will develop such a compiler algorithm. 

\item {\bf Extension to sparse tensors}. While there are many (more than 50 in 2010~\cite{Arnold:2010:SVS:1863543.1863581}) sparse representations for a sparse matrix, the possible representations for a sparse tensor can be overwhelming, since a sparse tensor can have many dimensions and any dimension can be dense or sparse. If a sparse tensor could be expressed in a similar way as a sparse matrix and for values-driven computations, programmers would be able to productively explore the design space, including not only the possible sparse tensor representations, but also their optimizations, for high performance.

\item {\bf More transformations}. While we have pointed out three high-level transformations, including packing, blocking, and job scheduling, it is possible to include more transformations to succinctly and precisely express the representations and optimizations of sparse matrices for various computations. There are known data parallel primitives that emerge in many computations~\cite{Bell:multigrid:2012,Sengupta:2007:SPG:1280094.1280110}, including reduction, scan (parallel prefix-sum), mapping, gathering, scattering, stream compaction, segmented reduction, and sorting~\cite{Bell:multigrid:2012}.  They might be used as additional high-level transformations. 

In this paper, the result of job scheduling is statically decidable with a sparse matrix. However, for GPUs and FPGAs, dynamic job scheduling is also common. It is useful if we could standardize dynamic job scheduling as a high-level transformation. 

\item{\bf Optimizing across computations}. 
So far, we have focused on a single sparse matrix computation. In an iterative solver, there are multiple computations involved. Reordering~\cite{Cuthill:1969:CM:800195.805928,George:1981:RCM:578296} may improve data locality. Inspection/execution is a common strategy that examines the structures of sparse matrices, and transforms the computations at run-time for better performance. In our previous work~\cite{Rong:2016:Sparso:2967938.2967943}, we have built a compiler and library to transparently optimize across computations with reordering and inspection/execution. Now that we can explicitly express a sparse representation, in future, we may leverage our previous work but further give programmers explicit control of reordering and inspection/execution.       

\item {\bf More workloads and designs for FPGAs}. In this paper, we have studied two SpMV implementations on FPGAs, and from them extracted a general approach. We will study more workloads and their designs on FPGAs to validate and enrich this approach. 

\item {\bf Workloads and designs for GPUs and CGRAs}. Similar to FPGAs, GPUs have massive resources and parallelism. CGRA (Coarse-Grain Reconfigurable Architecture)~\cite{Wijtvliet:CGRA:survey:2016} is structurally similar to FPGAs except its resources have coarse granularity. It is interesting to see if our approach is applicable to workloads and designs on these architectures.

\end{itemize}

\section{Conclusion}
\label{sec:conclusion}

We have proposed a novel idea to productively expressing a sparse matrix for high-performance computing on a spatial architecture. We express the sparse representation of the matrix as a series of invertible transformations, and express the computation centering around values, aided by an inverse mapping automatically generated by a compiler. Then a programmer can specify optimizations of the computation in a way much like for a dense matrix computation.

This is a first step in addressing spatial programming of sparse matrix computations for productive performance. There are many interesting  researches left for future.

\begin{acks} 
The keen interest of Nitish Kumar Srivastava in sparse tensors has motivated me to think about the problem in this paper (productive, performant, spatial programming of sparse matrix computations).  Zhiru Zhang, Nitish Kumar Srivastava, and Chris Hughes have provided valuable feedback on the solution.
\end{acks}

\bibliographystyle{abbrv}
\bibliography{local}

\end{document}